\documentclass[aps,prl,superscriptaddress,longbibliography,citeautoscript,twocolumn,10pt]{revtex4-2}
\usepackage{dutchcal}
\usepackage{float}
\usepackage{mathtools}
\usepackage{graphicx}%
\usepackage{bm}%
\usepackage[usenames,dvipsnames]{xcolor}
\usepackage[ colorlinks = true,
             linkcolor = violet,
             urlcolor  = violet,
             citecolor = ForestGreen,
             anchorcolor = ForestGreen,
]{hyperref}
\usepackage[normalem]{ulem}

\newcommand{\ket}[1]{|#1\rangle}
\newcommand{\bra}[1]{\langle #1|}

\newcommand{\Tr}{\mathrm{Tr}}

\newcommand{\st}[1]{\emph{#1.}---}

\begin{document}

\title{Protecting Quantum States via the Super-Zeno Effect and Anticoherence}

\affiliation{School of Physics and Optoelectronic Engineering, Hainan
  University, Haikou, 570228 People’s Republic of China }

\author{C.{} Chryssomalakos} 
\email{chryss@nucleares.unam.mx}
\affiliation{Instituto de Ciencias Nucleares, Universidad Nacional Aut\'onoma de
  M\'exico, PO Box 70-543, 04510, Ciudad de M\'exico, M\'exico}

\author{A.{} G.{} Flores-Delgado} 
\email{ana.flores@correo.nucleares.unam.mx}
\affiliation{Instituto de Ciencias Nucleares, Universidad Nacional Aut\'onoma de
  M\'exico, PO Box 70-543, 04510, Ciudad de M\'exico, M\'exico}

\author{E.{} Guzm\'an-Gonz\'alez} 
\email[Corresponding author: ]
{edgar.guzman@hainanu.edu.cn} \affiliation{School of Physics and Optoelectronic
  Engineering, Hainan University, Haikou, 570228 People’s Republic of China }

\author{L.{} Hanotel} 
\email{khanotel@hse.ru} 
\affiliation{Tikhonov Moscow Institute of Electronics and Mathematics, HSE University, Moscow 123458, Russia}

\date{\today}

\begin{abstract}
  We consider a spin system coupled to a noisy environment via an interaction Hamiltonian $H$ which is a polynomial of degree $n$ in the spin operators $S_i$ ($n$-\emph{magnetic} Hamiltonian), and show that the use of the quantum super-Zeno effect in conjunction with a spin-anticoherent subspace, is highly effective in protecting the spin from the noise. When $n=1$, as in the case of unwanted rotations induced by stray magnetic fields, our scheme provides a quantum gyroscope, a concept that generalizes naturally to higher $n$ values (a quantum $n$-gyroscope). The gist of the proposed protocol is to use the quantum super-Zeno effect to confine
  a state to the subspace, and then exploit anticoherence to freeze its evolution within it.
\end{abstract}

\maketitle

Quantum computing
exploits quantum effects to process and manipulate information more efficiently
than its classical counterpart. However, quantum states are inherently fragile, making
error correction essential for reliable computations. A common approach is to
embed the information-carrying quantum states in a linear subspace of a quantum
system that includes additional degrees of freedom, introducing redundancies
that enable error detection and correction \cite{Nielsen_2012}. If the error rate per operation remains below a certain
threshold, quantum error correction can suppress errors effectively, enabling
arbitrarily long quantum computations \cite{Knill_1998}. Since these errors
arise from imperfections in gate implementations and from perturbations in the
system caused by the environment, mitigating noise is of paramount importance in quantum computing.

One approach to achieving the above involves mapping the linear
subspace to be protected onto one that is naturally resistant to environmental noise, then allow the interaction to take place, and,
finally, mapping it back at the end of the interaction. A standard example is a
\emph{Decoherence-Free subspace}, which is a subspace of the system's Hilbert
space that remains invariant under the system's internal evolution, and where the
interaction Hamiltonian acts trivially, thereby providing perfect
protection \cite{Zanardi_1997,Duan_1997,Lidar_1999,Bei.Bra.Tre.Kni:00,Zan.Cam:15,Lidar_2003}. However, such
subspaces do not always exist.
In particular, if the system–environment interaction leaves no non-trivial subspace  invariant, Decoherence-Free subspaces cannot be found.

A second approach involves the  quantum Zeno effect
\cite{Facchi_2002,Erez_2004,Brion_2005,Fac.Nak.Pas.Tas:05,Paz_Silva_2012,Pop.Ess.Pre.Sch:18,Dominy_2013}, in which the system  is frequently
measured, while it interacts with the environment, to determine whether it remains within a particular subspace, resulting in the
system's dynamics being constrained to this subspace with high probability
\cite{Misra_1977,Facchi_2008}. As the frequency of measurements increases, the
\emph{survival probability} of the state inside the subspace
approaches one, making the subspace, as a whole, essentially invariant under noise.
Additional correction protocols can be applied to control the evolution of the
system within the subspace.

The high-frequency measurements necessary in the quantum Zeno scheme often present a challenge in practice. An alternative approach uses the so-called super-Zeno effect, in which unitary pulses are applied at judiciously chosen time intervals,
to constrain the system's evolution to the desired subspace. In this way, a higher
survival probability is, in principle, possible with the same number of interventions (compared to the standard Zeno approach), while maintaining the evolution unitary ---these theoretical predictions have been
experimentally demonstrated in the last couple of decades~\cite{Ting_Ting_2009,Singh_2014}.

We consider here the application of the super-Zeno effect to
the protection of a spin state from its interaction with a noisy environment ---the novel element of our approach is the use of spin anticoherent subspaces \cite{Pereira_2017}, which we define formally below. The interaction Hamiltonian is assumed to be a polynomial of degree $n$ in the spin operators $S_i$ (an \emph{$n$-magnetic Hamiltonian})~
\footnote{%
The use of the Lie algebra relations $S_i S_j-S_j S_i=i \epsilon_{ijk}S_k$ allows for changes in the polynomial order alluded to in the text --- the latter becomes well-defined once a particular ordering, \emph{e.g.}, $S_x^m S_y^r S_z^p$, of the factors in a monomial is adopted. This subtlety is dealt with formally in the Poincar\'e-Birkhoff-Witt theorem~\cite{Hall2003}.
}\nocite{Hall2003},
and is averaged over all orientations to model its stochastic nature. Our
motivation is twofold: (i)
upon identifying spin states with permutation-symmetric multiqubit states
(see \cite{SM}),
\nocite{Hebenstreit_2022, Wieczorek_2009, Wang_2018, Singh_2022, Toth_2007, Aulbach_2012, Devi_2011, Viola_2001, Fortunato_2002,  Serrano_2025, Bag.etal:15, Chryssomalakos_2022, Stanley2012, Bie.Doh.Uys:11, Sou.Alv.Sut:12}
this interaction arises naturally from symmetric system–environment coupling~\cite{Knill_1997,Knill_2000}, and (ii) since, in general,
there is no subspace of the state space of a spin invariant under an arbitrary
interaction of this type,
Decoherence-Free subspaces cannot be used in a straightforward way, making it a
simple example where the application of alternative protocols is necessary. Key
questions that we address include: (i) What is the effect of anticoherence on the efficiency of the super-Zeno scheme? (ii)  How does the fidelity of a protected state typically decrease as a
function of time? (iii) How does the (anticoherent) super-Zeno effect compare with the regular Zeno effect? The main contribution of the present work consists in pointing out the relevance of the use of anticoherent subspaces in the super-Zeno effect, as it is shown that the protection afforded to spin states within such subspaces is far superior to that attainable with generic ones.

\st{Spin anticoherent subspaces}
We now consider the problem of identifying subspaces of the spin-$s$ state space
that are robust against general $n$-magnetic interaction Hamiltonians. Natural
candidates for this problem are spin anticoherent subspaces \cite{Pereira_2017},
a direct generalization of anticoherent states \cite{Zimba_2006}. Since anticoherent states
exhibit minimal directionality, \emph{i.e.}, vanishing spin
expectation values and isotropic higher moments, these subspaces
inherit a similar isotropy.

One (among several equivalent)
definitions is the following: a
 spin-$s$, $k$-dimensional $q$-anticoherent subspace $\mathcal P$ is such that, for every basis $\{\ket{\psi_i}\}_{i=1}^k$
of $\mathcal P$ and all $1 \le l \le q$, $\langle \psi_i | (\mathbf S \cdot \mathbf n)^l | \psi_j \rangle$
is independent of the direction of $\mathbf n$, where $\mathbf S=(S_x,S_y,S_z)$ denotes the spin vector operator.

We show now that anticoherent subspaces can
be used to partially protect states from an $n$-magnetic
Hamiltonian $H$. Consider a $q$-anticoherent subspace $\mathcal{P}$, with $q = Mn$ for some integer $M$, an $n$-magnetic Hamiltonian $H$ (so that $q$ is an integer multiple of $n$), and an initial state $\ket{\psi} \in \mathcal{P}$.
Then, the projection of the
time-evolved state $\ket{\psi(t)}=e^{-itH}\ket{\psi}$ onto $\mathcal{P}$ is given by,
\begin{align}
  P e^{-i H t} \ket{\psi} 
  &= 
  \Biggl(
  \frac{1}{2s+1} \sum_{j=0}^M \frac{(-i t)^j}{j!} \Tr( H^j )
  \Biggr)\!
  \ket{\psi} + \mathcal{O}(t^{M+1})
  \nonumber \\
  & \equiv
  p_M(t) \ket{\psi} + \mathcal{O}(t^{M+1})
  \, ,
  \label{eq:evolProjAnti}
\end{align}
which follows from $q$-anticoherence,
as the isotropy of $\mathcal P$ up to order $Mn$ ensures that $P H^j P \propto \Tr(H^j) P$ for all $j \le M$ \cite{SM}.
Hence, to order  $ t^M $, the component of $\ket{\psi(t)}$ along $\mathcal{P}$ is only multiplied by the scalar factor $p_M(t)$, which is a polynomial of degree $M$ in $t$, leaving the dynamics within
$\mathcal P$ effectively \emph{frozen}, to that order in $t$. The modulus of $p_M(t)$, which is related to
the survival probability, is in general smaller than one, indicating a finite probability of the state leaking into the orthogonal complement $\mathcal{Q} \equiv \mathcal{P}^\perp$, where
all nontrivial dynamics occurs. Under these conditions, the inner dynamics of $\mathcal P$ can be frozen to
any order in $t$ (in the above sense), provided the order of anticoherence is high enough. Note that the standard Taylor/Lagrange bound on the omitted terms on the right hand side of~(\ref{eq:evolProjAnti}) is $t^{M+1}s^{M+1}/(M+1)!$, while tighter bounds (and, correspondingly, wider ranges of validity) can be determined in terms of the upper incomplete gamma function, after applying the triangle inequality to the remainder (see, \emph{e.g.}, \S 6.5 in~\cite{Abr.Ste:64}).

\st{Super-Zeno effect}
The freezing guaranteed by an anticoherent subspace $\mathcal{P}$ would be enough to fully protect a state within it, if there were no leaks outside of $\mathcal{P}$. However, such leakage, namely into the orthogonal complement of $\mathcal{P}$, does occur, and this is precisely where the super-Zeno scheme becomes essential.
The standard quantum Zeno effect can constrain the evolution of a system initially prepared in $\mathcal{P}$ to remain within it, but only to order $t$ ---the survival probability
decreases as $t^2$ \cite{Facchi_2008}. This implies that the extra protection
provided by higher orders of anticoherence would be lost.
The super-Zeno effect, on the other hand, provides survival probabilities that decrease like
$t^{2m}$ for arbitrary $m$ \cite{Dhar_2006}. This is achieved by applying to the system $K=K(m)$ (or more) 
unitary pulses or ``kicks'' $J=Q-P$, where $Q$ denotes the projector onto the
orthogonal complement $\mathcal{Q}$, 
with the timing of the kicks given by intervals $t_i$, $i=1,\dots,K+1$, which sum to the total time $t$. Specifically, the evolution operator is
\begin{equation}
  \begin{split}
U = e^{-i H t_{K+1}} J \dots J e^{-i H t_2} J e^{-i H t_1},
  \end{split}
  \label{eq:USuperZeno}
\end{equation}
with $t_1$ the time before the first kick, $t_{K+1}$ the time after the last, and $t_2,\dots,t_K$ the intervals between consecutive kicks.

\begin{figure}[t]
	\centering
	\includegraphics{ 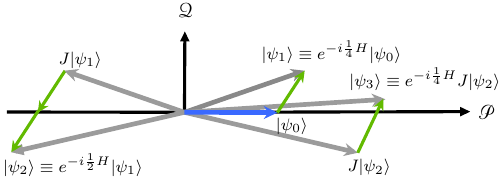 }
    \caption{Schematic illustration of the super-Zeno effect with two kicks, corresponding to the three time intervals $\{t_1,t_2,t_3\}=\{1/4,1/2,1/4\}$. The resulting evolution operator is given by $U=e^{-i\frac{1}{4}H}Je^{-i\frac{1}{2}H}Je^{-i\frac{1}{4}H}$, giving rise to the sequence of states $\ket{\psi_0} \rightarrow \ket{\psi_1} \rightarrow J\ket{\psi_1} \rightarrow \ket{\psi_2} \rightarrow J\ket{\psi_2} \rightarrow \ket{\psi_3}$, the last one returning (approximately) to the plane $\mathcal P$. State vectors are shown in grey, except for the first one which is shown in blue, and infinitesimal evolution vectors are in green. It is assumed in this figure (but not in the main text) that $\big|\ket{\psi_1}-\ket{\psi_0} \big| \ll 1$. Note also that the apparent rescaling of the state vectors is an artifact of the 2D nature of the sketch.}
    \label{fig:schematic}
\end{figure}

Geometrically, for $\ket{\psi}\in \mathcal P$ and $\ket{\phi}\in \mathcal Q$, the relations $J\ket{\psi}=-\ket{\psi}$ and $J\ket{\phi}=\ket{\phi}$ show that $J=Q-P$ acts as a reflection about the subspace $\mathcal Q$. The repeated action of such reflections reverses the leakage from $\mathcal P$, thereby constraining the dynamics; see Fig.~\ref{fig:schematic}.

As an example, the $m=5$ behavior can be achieved by taking $K=4$ kicks, and the following time intervals among them~\cite{Dhar_2006},
\begin{equation}
  \{t_1,t_2,t_3,t_4,t_5\}=
  \Bigl\{\beta t,\frac{t}{4} ,\frac{t}{2} (1-4 \beta),\frac{t}{4}, \beta t\Bigr\},
  \label{eq:t5Intervals}
\end{equation}
where $\beta =(3-\sqrt{5})/8$. Note that the above $t_i$ sum up to $t$, which can be as large as desired. The time intervals appropriate for attaining an $m=6$, $7$, behavior can be consulted in \cite{SM} (note that there is a typo in~\cite{Dhar_2006} in the $m=6$ intervals).

Since the super-Zeno effect confines the evolution of a state to $\mathcal{P}$ to arbitrary order in $t$, and the order of anticoherence controls how strongly the dynamics is frozen within $\mathcal{P}$, we expect that, by choosing a $\mathcal P$ with high enough order of anticoherence, the overall time evolution of a state can be frozen to any given order in $t$.

\st{Fidelity of a subspace}
A natural measure of the protection provided by the use of anticoherent subspaces  in the super-Zeno protocol is the fidelity $F$ between the initial state $\ket{\psi}$ and the evolved state $U \ket{\psi}$, averaged over all states in the subspace \cite{Pedersen_2007,Nielsen_2012},
\begin{equation}
  \begin{split}
F &= \int_{\mathcal{P}} d \psi \,|\!\bra{\psi} U \ket{\psi}\!|^2
    =
 \frac{
| \Tr(U_P) |^2 + \Tr(U_P  U^\dagger_P)
    }{k(k+1)},
  \end{split}
  \label{eq:AverageFidelity}
\end{equation}
where $U_P = P \, UP$ denotes the restriction of $U$ to $\mathcal P$, $k=\Tr(P)$ is the dimension of $\mathcal P$ and $d \psi$ denotes the intrinsic Fubini-Study measure on the subspace, normalized so that $\int d\psi=1$. Other choices of measure are possible, and probably more appropriate in specific contexts, but the essential points of our analysis seem to not depend on this choice, so we adopt the one mentioned above, mostly for simplicity.
Note that for any $\ket{\psi}\in \mathcal P$, $P\ket{\psi}=\ket{\psi}$. Hence, if $U_P \approx P$ (in an appropriate sense), then $U_P\ket{\psi}\approx\ket{\psi}$, so states in $\mathcal P$ remain almost static and are thus well-protected.

The specific form of $U_P$ depends on $H$, $P$, and
the choice of time intervals. To illustrate this, consider the $n=1$ case of
$H=\bm{B}\cdot\bm{S}$ and the five time intervals of Eq.~(\ref{eq:t5Intervals}).  We find $U_P = P+i \lambda_q PH^rPt^r+ \mathcal{O}(t^{r+2})$ where $r=3$
for $q=1$, $2$, $r=5$ for $q=3$, $4$, and so on, up to $r=9$ for $q\geq 7$, which is the maximum $r$-value that can be achieved with four kicks. For the above $U_P$  the fidelity $F$ decreases like 
$F = 1 - \mu_q(P)\, t^{6}$ for $q = 1, 2$, and as
$F = 1 - \mu_q(P)\, t^{10}$ for $q \geq 3$, with the first nontrivial power of $t$
depending only on $q$, and the choice of a particular $q$-anticoherent $\mathcal{P}$ affecting the result only through $\mu_q(P)$.
Note that the protection provided by $1$- and $2$-anticoherent subspaces
is of the same order in $t$,
necessitating the consideration of $3$-anticoherent subspaces to achieve a higher order of protection.
Some explicit computations of the constant $\mu_q(P)$ that appears above are provided in \cite{SM}.

It is remarkable that, in principle, with just four kicks, fidelities that decay
as $1-F\propto t^{10} $ can be achieved. The main theoretical challenge is
identifying subspaces with a high enough order of anticoherence, and there is little and fragmented knowledge about this in the literature. We note in passing that, currently, the highest order of
anticoherence reported for a two-dimensional subspace is
$5$, corresponding to spin $s = 30$ \cite{Pereira_2017}. However, with growing
interest in anticoherent spaces, more efficient algorithms for identifying them
are likely to emerge soon.

\st{Quantum gyroscopes}
The scheme presented above, specified to the $n=1$ case, $H^{(1)}=\mathbf{B}\cdot \mathbf{S}$,  can be used to protect a spin system from unwanted rotations, resembling the
functionality of a classical gyroscope. But our proposal can handle more general $n$-magnetic interaction Hamiltonians, realizing  \emph{quantum $n$-gyroscopes}. We illustrate this application in a
spin $s = 13/2$ system, the lowest spin value for which we have identified a
$3$-anticoherent $2$-plane \footnote{The analytical characterization of \(q\)-anticoherent \(k\)-dimensional subspaces for arbitrary spin \(s\) is, in general, a challenging problem, that is being currently pursued by several groups, including ours. However, such subspaces can be identified numerically, using, for example, non-unitary operators known as \(t\)-boosts \cite{Ara.etal:25}.}. The spin-13/2 projective space  serves as an ideal laboratory, in which the fidelities achieved by quantum $n$-gyroscopes, based on subspaces of various orders of anticoherence, may be numerically analyzed.

Consider the case in which an $n$-gyroscope is used to protect  the states in a two-dimensional subspace $\mathcal P_0$, from an interaction $H = (\bm{B} \cdot \bm{S})^n$,
where $\bm B$ is a stray  magnetic field, acting as noise. To this end, first, we map $\mathcal P_0$
to an anticoherent subspace $\mathcal P$ ---this \emph{coding map} can, in principle, be constructed for any pair of subspaces \cite{Brion_2005,Akulin_2001}.
Throughout the (total) interaction time $t$, we apply the operator $J$ every time step $t_i$ to
implement the operator $U$ in
Eq.~(\ref{eq:USuperZeno}). Finally, we apply the inverse coding map from $\mathcal P$
back to $\mathcal P_0$. We compute the resulting fidelity $F$ (recall that this already involves averaging over the subspace) and, since the magnetic field acts as noise, we further average $F$  over the direction of the field, yielding a doubly-averaged state fidelity $\bar{F}$. 

Assuming no additional errors occur when mapping $\mathcal P$ to
$\mathcal P_0$ and vice versa,
we study the  fidelity $\bar{F}$
for three different gyroscopes, using 2-dimensional subspaces of anticoherence $q=1, \, 2,\, 3$, respectively, spanned by states that are given explicitly in the Supplemental Material \cite{SM}.

\st{Numerical results}
In our numerical analysis we have found that averaging the fidelity over the direction of the magnetic field 
using
the vertices of an icosahedron (a spherical (2,2)-design~\cite{Hug.Wal:21}) yields an excellent approximation to results obtained by averaging over 100 points sampled uniformly on the sphere. Accordingly, the fidelity $\bar{F}$ depicted in the plots that follow has been obtained with this technique.

To measure how well the dynamics within the protected subspace $\mathcal{P}$ remains
``frozen'' during evolution, we compute the \emph{freezing fidelity} $F_{\mathcal{P}}$, defined as the fidelity  between a state $\ket{\psi} \in \mathcal{P}$
and the normalized projection of $U \ket{\psi}$ onto $\mathcal{P}$, averaged over all
states $\ket{\psi}$ in $\mathcal{P}$ ---the precise expression is
\begin{equation}
  F_{\mathcal{P}} = \int d\psi \,\frac{|\!\bra{\psi} P U \ket{\psi}\!|^2}{\bra{\psi} U^\dagger P U \ket{\psi}}
  \,.
  \label{eq:FreezingFidelityDef}
\end{equation}
When $\mathcal{P}$ is 2-dimensional, $F_{\mathcal{P}}$ can be computed in closed form ---the resulting expression is given in \cite{SM}.
Similarly to $\bar{F}$, we define $\bar{F}_{\mathcal{P}}$ as the average of
$F_{\mathcal{P}}$ over the direction of the magnetic field, and use the same icosahedral approximation in numerics. All simulations use $m=5$, corresponding to four super-Zeno kicks applied at the intervals defined in Eq. \eqref{eq:t5Intervals}, consistent with the order-three anticoherence of the anticoherent subspace considered and with current experimental implementations \cite{Singh_2014}. Finally, in the plots of fidelities \emph{vs} time we use  normalized Hamiltonians 
\begin{equation}
\label{normalization_plots}
\frac{H}{\sqrt{(2s+2)^{-1}(\text{Tr}(H^2)-(\text{Tr}H)^2/(2s+1))}}\ . 
\end{equation}

\begin{figure}[t]
	\centering
	\includegraphics[scale=0.435]{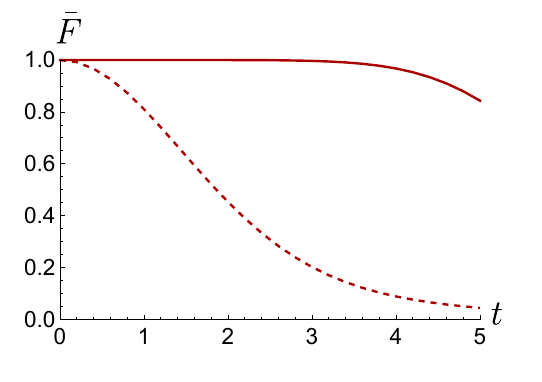}
    \hspace{3ex}
    \includegraphics[scale=0.435]{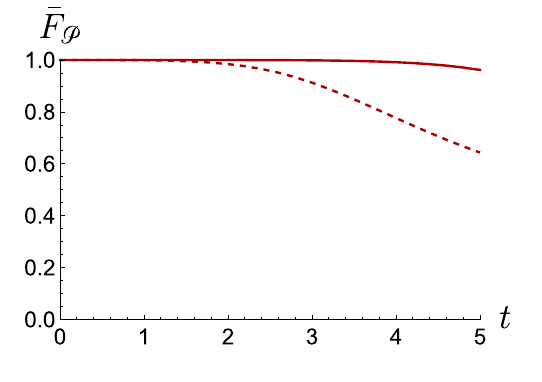}
	\caption{Left: Comparison of the (doubly) average fidelity $\bar{F}$ between the Zeno (dashed curve) and super-Zeno (solid curve) schemes for the spin-$13/2$ 1-anticoherent 2-plane given in~\cite{SM}. The interaction  Hamiltonian is $H^{(1)} \propto \mathbf{B} \cdot \mathbf{S}$,  
    normalized according to \eqref{normalization_plots}, and the corresponding fidelity is averaged by letting  $\mathbf{B}$ point along the vertices of an icosahedron. Both schemes involve the same number of operations ---four measurements for Zeno, four pulses for super-Zeno. Right: Similar comparison as on the left but for the freezing fidelity $\bar{F}_{\mathcal P}$.
    }
	\label{fig:comparisonZenoVSSuperZeno}
\end{figure}

Previous proposals used the regular Zeno effect, rather than the super-Zeno effect, to protect quantum states~\cite{Brion_2005} ---in Fig.~\ref{fig:comparisonZenoVSSuperZeno} (left frame) we compare the corresponding fidelities $\bar F(t)$ for gyroscopes involving the 1-anticoherent 2-plane given in~\cite{SM} ---the superiority of the super-Zeno is evident. A similar analysis for the freezing fidelity $\bar{F}_{\mathcal P}$ is presented in the right frame of the same figure. Because a high total fidelity requires both a large survival probability and effective freezing of the internal dynamics, the much smaller difference observed in $\bar{F}_{\mathcal P}$ shows that the enhancement in $\bar F(t)$ primarily arises from the higher survival probabilities provided by the super-Zeno effect, while internal freezing is largely determined by the order of anticoherence of the subspace, which is common to both protocols.

\begin{figure}[t]
\centering
    \includegraphics[scale=0.6]{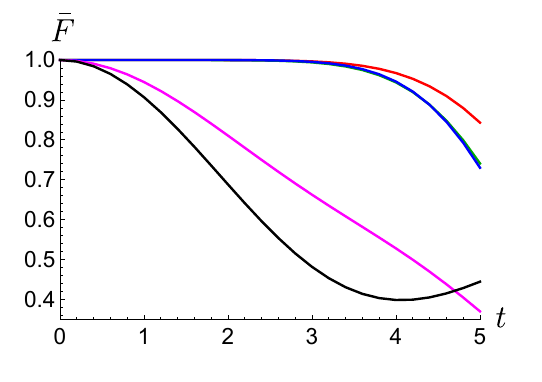}
\caption{Comparison of $\bar{F}$ for different orders of anticoherence, using 4 pulses and averaging $H^{(1)} \propto  \mathbf{B} \cdot \mathbf{S}$ over the direction of $\mathbf{B}$. The Hamiltonians were normalized as in \eqref{normalization_plots}.
The red, green (hidden behind the blue), and blue curves correspond to the 1-, 2-, and 3-anticoherent planes given in~\cite{SM}, respectively; the black curve corresponds to the coherent plane, and the magenta curve to a generic plane, as explained in the text.}
\label{fig:comparisonAnticoherence}
\end{figure}

Figure~\ref{fig:comparisonAnticoherence} shows the super-Zeno fidelity $\bar{F}(t)$ for subspaces of equal dimension and different anticoherence orders in the presence of the interaction Hamiltonian $H^{(1)}$.
$\bar{F}(t)$ behaves similarly for all anticoherent subspaces, but drops significantly when non-anticoherent subspaces are considered ---the latter were the coherent 2-plane spanned by the vectors
$\ket{13/2}$ and $\ket{11/2}$, which, naturally, can be considered the least anticoherent 2-plane available, and a generic plane that is neither coherent nor anticoherent, according to the polarization measure introduced in \cite{Hoz.Kli.etal:13}.

\begin{figure*}[t]
\centering
\includegraphics[scale=0.58]{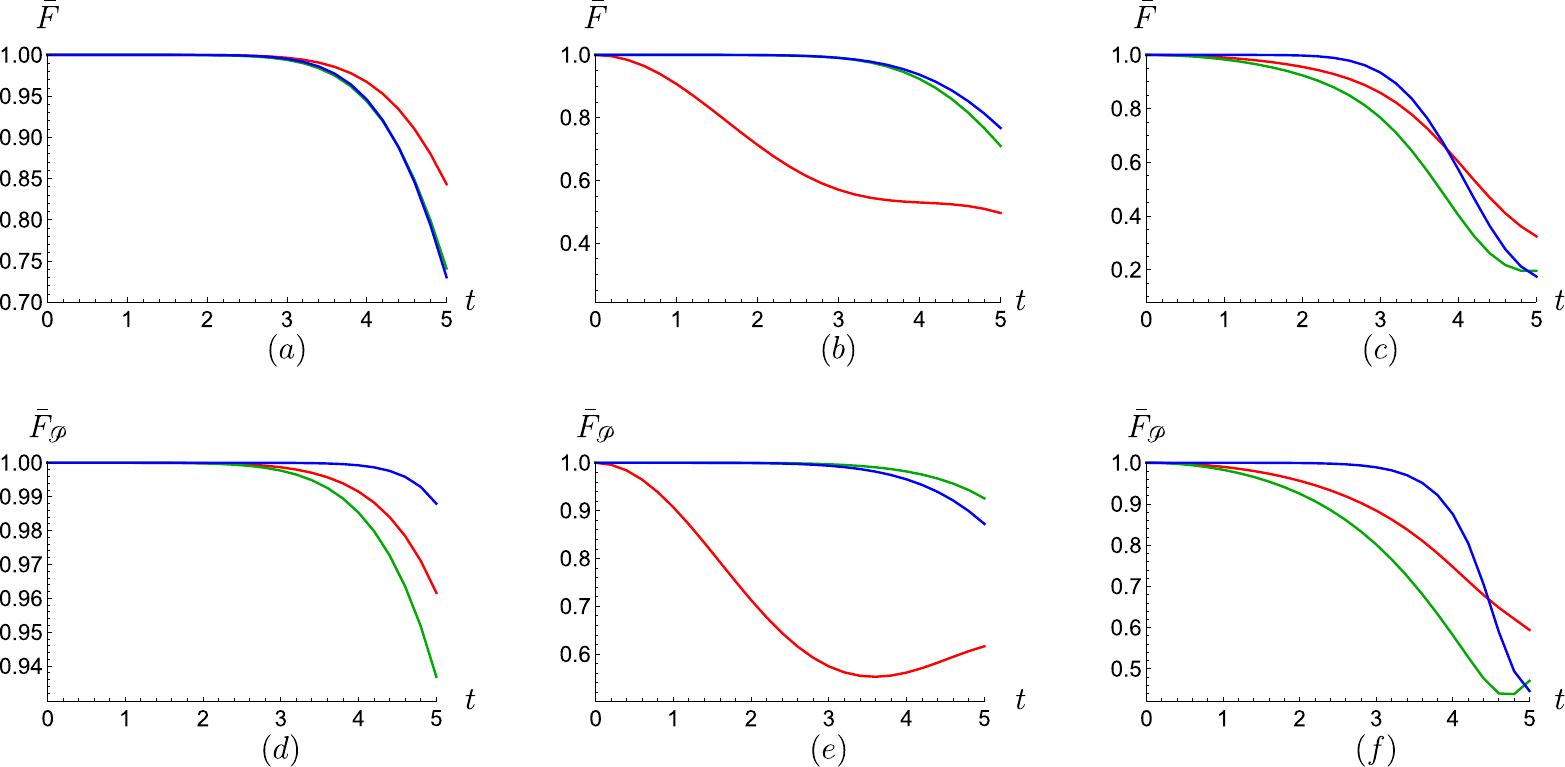}
\caption{Comparison of $\bar{F}$ (top row) and $\bar{F}_{\mathcal P}$ (bottom row) for the super-Zeno scheme with 4 pulses.
The red, green, and blue curves correspond to the 1-, 2-, and 3-anticoherent subspaces given in~\cite{SM}, respectively.
 The Hamiltonians considered are of the form $H \propto (\mathbf{B} \cdot \mathbf{S})^n$ and normalized according to \eqref{normalization_plots}, with $n=1$ (first column), $n=2$ (second column), and $n=3$ (third column). In plots (a), (b) and (e), the blue curve partially blocks the green one. Note also the difference in the vertical scales used: for $n=1$, the average fidelities are significantly higher, for the same value of $t$, than in the other cases.}
\label{fig:1SZ}
\end{figure*}

In Fig.~\ref{fig:1SZ}, the fidelities $\bar{F}(t)$ (top row) and $\bar{F}_{\mathcal{P}}(t)$ (bottom row) are plotted for gyroscopes
with various orders of anticoherence (red, green, and blue curves), corresponding to the 2-planes defined in the Supplemental Material \cite{SM}, and for different $n$-magnetic Hamiltonians ($n=1$ in the first column, $n=2$ in the second, and $n=3$ in the third).
\begin{figure*}[t]
\centering
\includegraphics[scale=0.58]{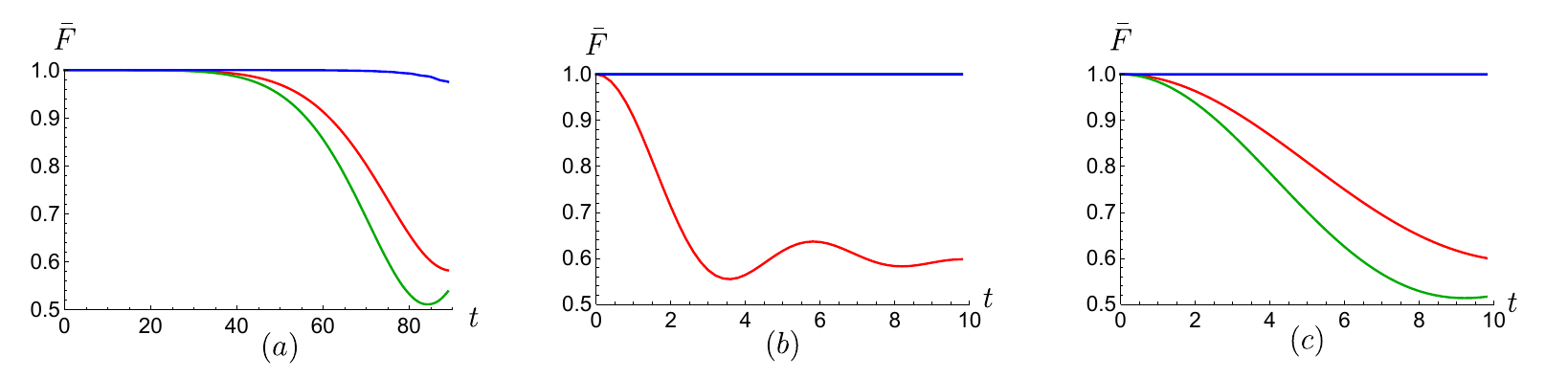}
\caption{
Comparison of $\bar{F}$ after applying the 4-pulse super-Zeno scheme 30 consecutive times (the total duration, in each case, can be read off the horizontal axis).
The red, green, and blue curves correspond to the 1-, 2-, and 3-anticoherent 2-planes given in~\cite{SM}, respectively.
Using the normalization of \eqref{normalization_plots}, we consider a Hamiltonian of the form $H \propto (\mathbf{B} \cdot \mathbf{S})^n$, with (a) $n = 1$, (b) $n = 2$, and (c) $n = 3$. In plot (b), only two curves are visible because the blue curve completely blocks the green one. Note also the difference in time scale in plot (a). In this grid, we omit the curves for $\bar{F}_{\mathcal{P}}$, since they are visually indistinguishable from those of $\bar{F}$, indicating that after 30 applications of the super-Zeno pulse sequence, the survival probability is so high that most of the fidelity loss arises from the evolution within the anticoherent subspace.}
\label{fig:30SZ}
\end{figure*}
In the case $n=1$, no significant differences are observed among the three subspaces for short times up to $t\approx 3$. For 
$n=2$, the 2- and 3-anticoherent subspaces have essentially the same behavior, while the 1-anticoherent subspace shows a much faster decay of fidelity. For
$n=3$, the three subspaces exhibit similar performance; however, the 3-anticoherent subspace appears to perform better at short times, as it is the only one whose fidelity remains close to 1 up to
$t\approx 3$.

To protect states for
longer durations $t$, we can apply the same pulse sequence a certain number
$\ell$ of times before mapping $\mathcal{P}$ back to  $\mathcal{P}_0$. As a result, the system evolution operator becomes $U^\ell$,
with $U$ defined as in Eq.~(\ref{eq:USuperZeno}),
but evaluated over a shorter time $t/\ell$.
In the experimental work reported in \cite{Singh_2014}, the pulse sequence defined by
the time intervals in Eq.~(\ref{eq:t5Intervals}) was applied
$\ell = 30$ times to constrain the evolution of a system to a 2-dimensional subspace.
Taking up this idea, Fig.~\ref{fig:30SZ} shows plots analogous to those in the first row of Fig.~\ref{fig:1SZ},
but with the difference that the super-Zeno pulse sequence in~(\ref{eq:t5Intervals}) has been applied 30 consecutive times. In the case $n = 1$, the short-time 
higher fidelities shown in Fig.~\ref{fig:1SZ} accumulate, resulting in the 3-anticoherent
subspace achieving substantially higher fidelities for longer durations compared to the other
subspaces. For $n = 2$, the $2$- and $3$-anticoherent subspaces have essentially the
same performance, while for $n=3$, the 3-anticoherent subspace maintains fidelity close to one up to
$t=10$, while the other subspaces exhibit visible decay from $t \approx 2$.
Since the spread in the
energies has been normalized to one, the natural period of the system is of the order of $2\pi$, so that, \emph{e.g.}, in Fig.~\ref{fig:30SZ}(a), even after about 14 cycles, the states in the
subspace remain practically unchanged.

\st{Concluding remarks}
We propose the use of spin-anticoherent subspaces in the super-Zeno protocol to isolate states from $n$-magnetic interactions with the environment, resulting in a quantum $n$-gyroscope ---the case $n=1$ corresponds to the standard gyroscope that protects from unwanted rotations. As an example, we show that (doubly) average fidelities (over all states in a subspace and all directions of the magnetic field) that decrease as $1- \bar{F} \propto t^{10}$ are within reach, using a 3-anticoherent 2-plane of spin 13/2. These results could prove useful in the protection of permutation-symmetric multiqubit states that are subject to symmetric magnetic noise. Further research could explore non-symmetric states and interactions, the protection of mixed states, and the (considerably harder) problem of time-varying interactions. Some remarks on the robustness of the protocol proposed here are given in the Supplemental Material \cite{SM}.

\st{Acknowledgments}
The authors acknowledge the valuable comments of the anonymous reviewers. CC and AGFD would like to acknowledge partial financial support from the CBF-2025-I-676 SECIHTI and DGAPA-UNAM IN112224 projects. 
AGFD acknowledges financial support from SECIHTI through a research assistantship. EG-G acknowledges support from the National Natural Science Foundation of China under Grant No. W2511080.

\onecolumngrid
\newpage
\makeatletter

\begin{center}
\textbf{Supplemental material for\\ ``Protecting Quantum States via the Super-Zeno Effect and Anticoherence''} \\[10pt]

C.{} Chryssomalakos$^{2}$, A.{} G.{} Flores-Delgado$^{2}$, E.{} Guzmán-González$^{1}$
and
L.{} Hanotel$^{3}$ \\

 \vspace{10pt}

\emph{$^1$
School of Physics and Optoelectronic Engineering, Hainan
  University, Haikou, 570228 People’s Republic of China
}\\
 \vspace{5pt}
\emph{$^2$
Instituto de Ciencias Nucleares, Universidad Nacional Aut\'onoma de
  M\'exico, PO Box 70-543, 04510, Ciudad de M\'exico, M\'exico
}\\
 \vspace{5pt}
\emph{$^3$
Tikhonov Moscow Institute of Electronics and Mathematics, HSE University, Moscow 123458, Russia
}\\
\end{center}

\vspace{10pt}

\makeatletter
\setcounter{secnumdepth}{3}
\makeatother

\setcounter{section}{0}
\setcounter{page}{1}
\renewcommand{\thesection}{S\arabic{section}}
\makeatletter

\twocolumngrid

In this Supplemental Material we present:
\begin{itemize}
  \item A review of basic facts about permutation-symmetric qubit states and their connection to spin states.
  \item An illustrative example of the mapping between $n$-qubit symmetric
        interactions in a system of qubits and $n$-order magnetic interactions in a spin system.
  \item The mathematical proof of the equivalence between different definitions of  anticoherent subspaces, some comments on what is currently known about them, and some details of the 2-planes used in the main text.
  \item Optimal time intervals for the quantum super-Zeno effect that yield survival
        probabilities that decay as $1-\alpha t^{2m}$ using $K = m - 1$ kicks, from $m=2$
        up to $m=7$.
  \item Explicit expressions for the first non-trivial order in $t$ of $PUP$ for a Hamiltonian of the form $H=\bm{B} \cdot \bm{S}$,
        different orders of anticoherence, and when the time intervals of
        Eq.~(3) of the main text are used to implement the super-Zeno effect. We also provide similar expressions for $F$.
  \item A closed-form expression for the freezing fidelity $F_{\mathcal P}$ when $\mathcal P$ is two-dimensional.
  \item Some remarks on the robustness of the proposed protocol.
\end{itemize}

\section{Permutation-symmetric qubit states and their spin representation}
The  states of an $N$-qubit system
that are invariant under permutations of the qubits have proven to be
crucial in the study of quantum information. They have been implemented
experimentally \cite{Wieczorek_2009,Wang_2018,Singh_2022},
and appear naturally when qubits cannot be manipulated
individually~\cite{Toth_2007,Aulbach_2012}. In addition to their experimental
significance, their algebraic structure allows a very simple mathematical
description: their state space
is isomorphic to that of a spin $s=N/2$ \cite{Devi_2011}, the isomorphism commuting with the natural $SU(2)$ action, corresponding to physical rotations. Under this
mapping, the spin operator $S_i$ for the whole system corresponds to the symmetric sum of
single-qubit operators, $\sum_m \sigma_i^m/2$, where $\sigma_i^m$ denotes the Pauli operator
$\sigma_i$ acting on the $m$-th qubit.

With the help of this mapping, it is possible to study the time evolution of a
symmetric state of qubits that couple symmetrically with their environment.
Experimentally, this is often the case when the correlation length of the
environmental coupling is longer than the qubit spacing --- for instance, in a low
temperature experiment where only long-wavelength excitations of the environment
survive \cite{Viola_2001,Fortunato_2002,Lidar_2003}. In this setting, an arbitrary
single-qubit interaction of the system with an external field,
$H^{(1)}_{\text{int}} \sim \sum_{m,i} B^i \sigma^m_i$, translates to an interaction $2\bm B \cdot \bm S$ of a spin-$s$ system with
a magnetic field $\bm B$. In the same way, the most general
symmetric $2$-qubit interaction,
$H^{(2)}_{\text{int}} \sim \sum_{m \neq p}\sum_{i ,j}B^{i j} \sigma^{m}_{i}\sigma^{ p}_{j}$, corresponds to a \emph{quadratic
magnetic} (or, \emph{2-magnetic}) interaction of the form $\sum_{i ,j}B^{i j} (2\{S_i,S_j\}-N \delta_{ij})$ of the
spin with an external field $\bm B$.
As illustrated in the following section, a symmetric interaction of $n$ qubits corresponds to an \emph{$n$-magnetic} interaction Hamiltonian in the spin representation.

\section{Correspondence between $n$-qubit interactions and $n$-order magnetic
interactions}
We show how $n$-qubit interactions are mapped to $n$-order magnetic interactions
in the spin language. For simplicity, we restrict the exposition to the case $n=4$. The structure
of the general case is the same, although the combinatorics involved are
more complicated.

The most general symmetric $4$-qubit interaction can be written as a linear combination of
terms of the form $ \sum\nolimits' \sigma_{i}^{m} \sigma_{j}^{o} \sigma_{h}^{p} \sigma_{l}^{u}$,
where the sum $ \sum\nolimits'$ runs over all the configuration of indices $m$, $o$, $p$ and $u$,
such that all of them are different.
Since this is the case, all the operators inside the sum
commute. Therefore, $\sum\nolimits' \sigma_{i}^{m} \sigma_{j}^{o} \sigma_{h}^{p} \sigma_{l}^{u}=
\sum\nolimits' \sigma_{j}^{m} \sigma_{i}^{o} \sigma_{h}^{p} \sigma_{l}^{u}$, and the same holds true
for any permutation of the tuple $(i,j,h,l)$. To make this symmetry explicit,
we introduce the following symbol,
\begin{equation}
  \begin{split}
\eta^{abcd}=
    \frac{1}{\mathcal A}
\begin{cases}
1  & \text{if ($a$,$b$,$c$,$d$) is a permutation of ($i$,$j$,$h$,$l$)}  \\
0  & \text{if it is not}
\end{cases},
  \end{split}
\end{equation}
where
$\mathcal A$ denotes
the number of permutations of the tuple ($i$,$j$,$h$,$l$)--- for instance,
if $i=j=h=l$, $\mathcal A=1$; if $i=j=h=x$ and $l=y$ then $\mathcal A=3$, and so on.
Thus
\begin{equation}
  \begin{split}
\sum\nolimits' \sigma_{i}^{m} \sigma_{j}^{o} \sigma_{h}^{p} \sigma_{l}^{u}=
 \sum\nolimits' \eta^{abcd}\sigma_{a}^{m} \sigma_{b}^{o} \sigma_{c}^{p} \sigma_{d}^{u}
  \end{split}
\end{equation}
where there is an implicit sum over contracted indices.
An important advantage of introducing $\eta$ is that the resulting expression
is symmetric under the interchange of the indices $m$, $o$, $p$, $u$, even if we do not assume they are all different.
For instance, if we change $m$ with $o$,
\begin{equation*}
  \begin{split}
\eta^{abcd}\sigma_{a}^{o} \sigma_{b}^{m} \sigma_{c}^{p} \sigma_{d}^{u}&=
\eta^{abcd}( \sigma_{b}^{m}\sigma_{a}^{o}+[\sigma_{a}^{o}, \sigma_{b}^{m}])   \sigma_{c}^{p} \sigma_{d}^{u}
\\&=
\eta^{bacd} \sigma_{b}^{m}\sigma_{a}^{o}   \sigma_{c}^{p} \sigma_{d}^{u}
=
\eta^{abcd} \sigma_{a}^{m}\sigma_{b}^{o}   \sigma_{c}^{p} \sigma_{d}^{u},
  \end{split}
\end{equation*}
where we used that $\eta^{bacd}$ is symmetric in $ab$ while $[\sigma_{a}^{o}, \sigma_{b}^{m}]$
is antisymmetric to obtain the second equality.

The next step, is to write the sum $\sum \nolimits'$ in terms of sums where some of the indices
$m$, $o$, $p$ and $u$ are forced to be the same,  while the others vary freely
without requiring them to be different. This can
be done systematically using Möbius inversion formula \cite{Stanley2012}.

Within this framework, after taking advantage of the permutational symmetry just mentioned,
 we rewrite the sum over the qubits as follows,
\begin{equation*}
  \begin{split}
 \sum_{\mathclap{m\neq o \neq p \neq u}}=\,\,\,
 \sum_{\mathclap{m o p u}}
+6 \mu_1 \sum_{\mathclap{\substack{m =o,\\pu}}}
+4 \mu_2 \sum_{\mathclap{\substack{m =o=p,\\u}}}
+3 \mu_3 \sum_{\mathclap{\substack{m =o,\\p=u}}}
+\mu_4 \sum_{\mathclap{m =o=p=u}},
  \end{split}
\end{equation*}
where
$\mu_1=-1, \mu_2=2, \mu_3=1, \mu_4=-6$
are the Möbius numbers for the partitions of a set with 4 elements, and $6$, $4$ and $3$ are combinatorial
factors that appear due the permutational symmetry of the summand in the indices $m$, $o$, $p$, $u$.

The first sum in the right hand side can be computed straightforwardly,
\begin{equation}
  \begin{split}
 \sum_{m o p u} \eta^{abcd} \sigma_{a}^{m} \sigma_{b}^{o} \sigma_{c}^{p} \sigma_{d}^{u}=
16 \eta^{abcd} S_{a} S_{b} S_{c} S_{d}\,.
  \end{split}
\end{equation}
To compute the second sum, we proceed as follows,
\begin{equation*}
  \begin{split}
- 6 \sum_{m  p u} \eta^{abcd} \sigma_{a}^{m} \sigma_{b}^{m} \sigma_{c}^{p} \sigma_{d}^{u}
    &=-12\eta^{abcd} \sum_{m}  \{\sigma_{a}^{m},\sigma_{b}^{m}\} S_c S_d
      \\&
    =-24\eta^{abcd}N \delta_{ab} S_c S_d,
  \end{split}
\end{equation*}
where we exploited the permutational symmetry to write the anticommutator $\{\sigma_{a}^{m},\sigma_{b}^{m}\}/2$ instead of
$\sigma_{a}^{m}\sigma_{b}^{m}$, and use the fact that the Pauli matrices constitute a representation of the Clifford algebra,
$\{\sigma_{a}^{m},\sigma_{b}^{m}\}=2  \delta_{ab}$.

Similarly, we compute the remaining terms,
\begin{equation}
  \begin{split}
8 \sum_{m, u} \eta^{abcd}  \sigma_{a}^{m} \sigma_{b}^{m} \sigma_{c}^{m} \sigma_{d}^{u}
&=32 \delta_{bc} \eta^{abcd}   S_a S_{d},
  \\
3 \sum_{m, p} \eta^{abcd}  \sigma_{a}^{m} \sigma_{b}^{m} \sigma_{c}^{p} \sigma_{d}^{p}
    &=
    3 N^2 \eta^{abcd}  \delta_{ab} \delta_{cd},
  \\
-6 \sum_{m} \eta^{abcd}  \sigma_{a}^{m} \sigma_{b}^{m} \sigma_{c}^{m} \sigma_{d}^{m}
&=-6N \eta^{abcd} \delta_{ab}\delta_{cd}.
  \end{split}
\end{equation}
Therefore,
\begin{equation*}
  \begin{split}
\sum\nolimits' \sigma_{i}^{m} \sigma_{j}^{o} \sigma_{h}^{p} \sigma_{l}^{u}
    &=\eta^{abcd} (16S_a S_b S_c S_d
    -24N \delta_{ab} S_c S_d
      \\&{}+
32 \delta_{bc}    S_a S_{d}
    + 3N^2 \delta_{ab} \delta_{cd}
-6N  \delta_{ab}\delta_{cd}).
  \end{split}
\end{equation*}
So $\sum\nolimits' \sigma_{i}^{m} \sigma_{j}^{o} \sigma_{h}^{p} \sigma_{l}^{u}$ corresponds to a $4$-degree magnetic
interaction, as claimed. A similar procedure can be followed for any symmetric $n$-qubit interactions.

\section{Spin anticoherent subspaces: characterization and relevant examples}
An equivalent definition of an anticoherent subspace is the following.
A subspace $\mathcal P$ is anticoherent of order $q$ ($q$-anticoherent for short) if, for any $q$-magnetic Hamiltonian $A$,
it holds that
$\bra{\psi} R^\dagger A R \ket{\psi} = \bra{\psi} A \ket{\psi}$
for all $\ket{\psi}$ in $\mathcal P$ and all $SU(2)$ transformations $R$,
\emph{i.e.}, a rotation of $A$ cannot be detected using states in $\mathcal P$.
This equivalence follows from the fact that any
$q$-magnetic Hamiltonian $A$ can be written as a linear combination of terms of the form
$(\mathbf{S}\cdot \mathbf{n})^k$ with $k\le q$.

Using this definition, we now prove another equivalent characterization,
$\mathcal P$ is $q$-anticoherent if and only if
\[
P A P = \frac{\Tr(A)}{2s+1} \, P
\]
for all $q$-magnetic $A$.

Consider a $q$-magnetic $A$ and suppose that $\mathcal P$ is $q$-anticoherent.
Take any spin state $\ket{\psi}$, not necessarily in $\mathcal P$. Since $P \ket{\psi}$ lies in $\mathcal P$, it
follows from the definition of anticoherence that $\bra{\psi}P R^{\dagger} A R P
\ket{\psi}$ is independent of $R$. By averaging over all rotations using the Haar measure, we conclude
$\bra{\psi}P A
P \ket{\psi} = \bra{\psi}P \langle R^{\dagger} A R\rangle_R P \ket{\psi}$.
Notice that the operator $\langle R^{\dagger} A R \rangle_R$ is rotationally invariant by
construction. Since the group of rotations acts irreducibly on the state space
of a spin, this operator is a scalar multiple of the identity. Therefore,
$\langle R^{\dagger} A R \rangle_R= \bm{1}\times\Tr(A)/(2s+1)$.
Thus,
\begin{equation}
\bra{\psi}P A P \ket{\psi} = \bra{\psi} \Bigl(\Tr(A)P/(2s+1) \Bigr) \ket{\psi}.
\end{equation}
Since this equality holds for any state $\ket{\psi}$,
$P A P = \Tr(A)P/(2s+1)$ follows.

Conversely, take a rotation $R$, a state $\ket{\psi}$ in $\mathcal P$, and a
polynomial $A$ of degree $q$ in the spin operators. Note that $R^\dagger A R$ is
also a polynomial of the same type. By hypothesis,
$PR^\dagger A RP= \Tr(R^\dagger A R)P/(2s+1)=\Tr(A)P/(2s+1)$.
Therefore,
\begin{equation}
\bra{\psi} R^\dagger A R \ket{\psi} = \bra{\psi} P R^\dagger A R P \ket{\psi} = \frac{\Tr(A)}{2s+1}\,,
\end{equation}
so $\bra{\psi} R^\dagger A R \ket{\psi}$ is independent of $R$, and thus
$\mathcal P$ is $q$-anticoherent.

Several techniques have emerged to find anticoherent subspaces for various values of $s$
\cite{Pereira_2017,Serrano_2025}, but little is known in general about the subvariety of $q$-anticoherent $k$-subspaces in the spin-$s$ Hilbert space.
Although it is not known in general how the maximal achievable $q$ scales with $s$ for an arbitrary $k$, the case of states ($k=1$) is well understood \cite{Bag.etal:15}, and, for $q=1,2$, the relation between the dimension of the subspace and the spin has been analyzed in \cite{Serrano_2025}. What makes the problem of identifying such subspaces even more compelling is that they are a key ingredient of a protocol (\emph{toponomic quantum computation}) to produce noise tolerant quantum gates~\cite{Chryssomalakos_2022}, and they also appear naturally in mixed-state quantum metrology~\cite{Serrano_2025}.

The 2-planes used in the main text are spanned by
\begin{align}
 \ket{\psi^{(1)}_1}
 &=
 \ket{-13/2}+\ket{13/2}
 \, ,
\nonumber\\
\ket{\psi^{(1)}_2}
&=
\ket{-9/2} + \ket{9/2}
\, ,
\nonumber\\
\ket{\psi^{(2)}_1}
&=
\sqrt{13} \, \ket{-5/2}+\sqrt{5} \, \ket{13/2}
\, ,
 \label{psispanq}
\\
\ket{\psi^{(2)}_2}
&=
-\frac{\ket{-11/2}}{\sqrt{3}}-\frac{\ket{1/2}}{\sqrt{6}}+\frac{\ket{7/2}}{\sqrt{2}}
\, ,
\nonumber\\
\ket{\psi^{(3)}_1}
&=
\frac{\ket{-13/2}}{\sqrt{13}} + \frac{i \gamma \ket{-5/2}}{\sqrt{5}}
    - \frac{\gamma \, \ket{3/2}}{\sqrt{7}} - \frac{i \ket{11/2}}{\sqrt{7}}
\, ,
\nonumber\\
\ket{\psi^{(3)}_2}
&=
\frac{ \ket{-11/2}}{\sqrt{7}} - \frac{i\gamma \, \ket{-3/2}}{\sqrt{7}}
    - \frac{ \gamma \, \ket{5/2}}{\sqrt{5}} + \frac{i \ket{13/2}}{\sqrt{13}}
    \, ,
    \nonumber
   \end{align}
where $\ket{\psi^{(q)}_{1,2}}$ span the $q$-anticoherent 2-plane of spin $13/2$, $\ket{m}$ denotes the eigenstate of $S_z$ with eigenvalue $m$, and $\gamma \equiv (\sqrt{7}+2 i)/\sqrt{11}$.

\section{Optimal time intervals for the quantum super-Zeno effect}
By applying a certain number $K$ of pulses, the quantum super-Zeno effect can be used to confine a system to
a subspace with a survival probability that decays like $t^{2m}$
\cite{Dhar_2006,Ting_Ting_2009}. For small values of $m$, it is possible
to find a sequence of time intervals involving only $K=m-1$ pulses,
whose explicit forms are given below.

For $m=2$, we can take the time intervals,
\begin{equation}
  \begin{split}
\{t_1,t_2\}= \Bigl\{ \frac{t}{2},\frac{t}{2} \Bigr\}\,.
  \end{split}
\end{equation}
For $m=3$, we can take,
\begin{equation}
  \begin{split}
\{t_1,t_2,t_3\}= \Bigl\{ \frac{t}{4}, \frac{t}{2},\frac{t}{4} \Bigr\}\,,
  \end{split}
\end{equation}
while, for $m=4$,
\begin{equation*}
  \begin{split}
\{t_1,t_2,t_3,t_4\}= \Bigl\{ \frac{t}{2} \Bigl(1-\frac{1}{\sqrt{2}}\Bigr), \frac{t}{2\sqrt{2}},\frac{t}{2\sqrt{2}},
\frac{t}{2} \Bigl(1-\frac{1}{\sqrt{2}}\Bigr)\Bigr\}\,.
  \end{split}
\end{equation*}
For $m=5$, the time intervals are those of equation (3) of the main text.

For $m=6$, the required time intervals are (there is a typo in \cite{Dhar_2006}),
\begin{equation}
  \{t_1, \dots ,t_6\}
  = \Bigl\{ \frac{t}{4}\kappa, \frac{t}{4} (1-\kappa), \frac{t}{4}, \frac{t}{4}, \frac{t}{4} (1-\kappa), \frac{t}{4} \kappa \Bigr\},
\end{equation}
where $\kappa=2-\sqrt{3}$.

For $m=7$, seven time intervals are required. The first four are,
\begin{equation}
  \begin{split}
    \{t_1, \dots ,t_4\}=
    \Bigl\{
    \frac{t}{4} (1-2 \nu ),
    \frac{t}{2} \lambda,
    \frac{t}{2}\nu,
    \frac{t}{2} ( 1-2\lambda )
 \Bigr\},
  \end{split}
\end{equation}
where,
\begin{equation}
  \begin{split}
    \nu &=\frac{1}{2} \sqrt{\frac{7}{3}} \sin \theta+\frac{\sqrt{7} }{6} \cos \theta
-\frac{1}{3}\, ,
    \\
\lambda
&=
\frac{1}{2} \sqrt{\frac{7}{3}} \sin \theta
  -
    \frac{\sqrt{7}}{6}  \cos \theta
+\frac{1}{3}
\, ,
  \end{split}
\end{equation}
with $\theta=\arctan (3 \sqrt{3})/3$. The remaining times are obtained by reversing the order of the first three,
$\{t_5,t_6,t_7\}=\{t_3,t_2,t_1\}$.

For $m=8$, the application of eight or more pulses is required.

\section{Leading order for the fidelity for four pulses}
In this section, we give explicit expressions showing how the operator $U_P=PUP$
depends on the order of anticoherence $q$ of $\mathcal P$ when $H$ is of the
form $H=\bm B \cdot \bm S$ and the four pulses given by equation
(3) of the main text are applied.

By considering the definition (2) of the main text with the time intervals of
equation (3) of the main text and the property $PH^d P=\Tr(H^d) P / (2s+1)$,
valid for $d\leq q$, it is possible to show that $U_P$ takes the following form,
$U_P = P+i \lambda_q PH^{r_q}Pt^{r_q}+ \mathcal{O}(t^{r_q+2})$ where $r_q=3$
for $q=1,2$; $r_q=5$ for $q=3,4$ and so on, until $r_q=9$ for $q=7,8$.
The explicit expression for different values of $\lambda_q$ are,
\begin{equation}
  \begin{split}
\lambda_{1,2}=&\frac{1}{48}(3 \sqrt{5}-7),\\
\lambda_{3,4}=&\frac{1}{15360} (122-55 \sqrt{5}),\\
\lambda_{5,6}=&\frac{707 \sqrt{5}-1592}{2^{15}\cdot  3^2\cdot 5 \cdot 7},\\
\lambda_{7,8}=&\frac{17467-8079 \sqrt{5}} { 2^{22} \cdot 3^4 \cdot 5 \cdot 7}.
  \end{split}
\end{equation}
If we compute the next order of $PUP$, we can compute the first non-trivial order of
the fidelity $F$ of Eq.~(4) of the main text. As some illustrative
examples,  we give explicit expressions for $q=1,2$ and $q=3,4$.

For $q=1,2$, the result is,
\begin{equation}
  \begin{split}
F &=
    1+
\frac{47 - 21 \sqrt{5}}{1152k(k+1)}\Bigl([\Tr(PH^3P)]^2-k \Tr[(PH^3P)^2]
    \Bigr)t^6
\\& {}+\mathcal O (t^7).
  \end{split}
\end{equation}
where we recall that $k=\Tr(P)$ denotes the dimension of $\mathcal P$.

For $q=3,4$, the expression is considerably more complicated,
so we introduce new notation.
Given any operator $O$, define $O_P\equiv POP$. We will also
write $O^n_{\phantom{n}P}\equiv PO^n P$, that is different from $(O_P)^n$.
The fidelity $F$ is,
\begin{equation}
  \begin{split}
F &=
1+\frac{2(k+1) a_2+ a_1^2+\Tr(A_1^2)}{k(k+1)} t^{10} +\mathcal O (t^{11}),
  \end{split}
\end{equation}
where, $a_i = \Tr(A_i)$,
\begin{equation}
  \begin{split}
A_1=&
 \frac{ \left(122-55 \sqrt{5}\right) H^5_{\phantom{n}P}}{15360}
\,,\\
A_2=&-\frac{1}{2^ {23} \cdot 15^2 }
    \Bigl(
25 H^{10}_{\phantom{n}P} -
100 (5 - 2 \sqrt{5}) H^2_{\phantom{n}P} H^8_{\phantom{n}P}
\\&\qquad\qquad{}+
25(75 - 34 \sqrt{5}) \{H^4_{\phantom{n}P}, H^6_{\phantom{n}P}\}
\\&\qquad\qquad {}+
(120011 - 53680 \sqrt{5}) (H^5_{\phantom{n}P})^2
\\&\qquad\qquad {}+
100 (12 - 5 \sqrt{5})(H^2_{\phantom{n}P})^2 H^6_{\phantom{n}P}
\\&\qquad\qquad {}+
125 (1709 - 764 \sqrt{5})H^2_{\phantom{n}P} (H^4_{\phantom{n}P})^2
\\&\qquad\qquad {}
- 1500 (293 - 131 \sqrt{5}) (H^2_{\phantom{n}P})^3 H^4_{\phantom{n}P}
\\&\qquad\qquad {}
+1800 (123 - 55 \sqrt{5}) (H^2_{\phantom{n}P})^5
    \Bigr),
  \end{split}
\end{equation}
and as usual, $\{,\}$ denotes the anticommutator.

Increasing the protection beyond the results presented here and shown in the figures of the main text would require a larger number of super-Zeno kicks and states with higher order of anticoherence, both of which entail additional experimental resources such as tighter pulse control and more demanding state preparation. A detailed optimization of this trade-off between achievable fidelity improvement and experimental cost is beyond the scope of this Letter.
\section{Freezing fidelity for two-dimensional subspaces}
In what follows, we derive a closed expression for the freezing fidelity
$F_{\mathcal P}$ defined in equation (6) of the main text when
$\mathcal P$ is two-dimensional.

First, we express $F_{\mathcal P}$ using the operator $U_P$ as follows,
\begin{equation}
  \begin{split}
F_{\mathcal P} &= \int d \psi \,\frac{|\!\bra{\psi}  U_P \ket{\psi}\!|^2}{\bra{\psi} U_P^\dagger  U_P \ket{\psi}}\,.
  \end{split}
  \label{eq:FreezingFPUP}
\end{equation}
By construction, $U_P$  acts exclusively within $\mathcal P$.
If we write $U_P$ as a block matrix with components  corresponding to $\mathcal P$ and
 $\mathcal Q$,  only the component $\mathcal P \mathcal P$ is nonzero.
For simplicity, we identify $U_P$ with this block component throughout this section.

Since $\mathcal P$ is two-dimensional, after choosing an orthonormal basis of
$\mathcal P$, we can write $U_P$ as a complex linear combination of the Pauli
matrices $\sigma_x, \sigma_y, \sigma_z$ and the identity matrix $\bm 1$. Due to the invariance
of $F_{\mathcal P}$ under complex rescalings and under unitary transformations
of $U_P$, we can simplify the calculation by initially assuming the
following form for $U_P$,
\begin{equation}
\begin{split}
U_P = \bm{1}+ (x_R+i x_I)\sigma_x -i \cot w \,\sigma_y,
\end{split}
\label{eq:upSimple}
\end{equation}
where $w \in (0, \pi)$, and $x_R, x_I$ are real. At the end of the calculation, we will rewrite the
result in terms of quantities invariant under these transformations,
making it valid for arbitrary $U_P$.

Next, we compute the eigenstates of $U_P^\dagger U_P$. For
the simple form given in (\ref{eq:upSimple}), a straightforward computation yields the following eigenbasis,
\begin{equation}
\begin{split}
\ket{\psi_1}&= \bigl(\cos(w/2), \sin(w/2)\bigr) , \\
\ket{\psi_2}&= \bigl(\sin(w/2),- \cos(w/2)\bigr) .
\end{split}
\end{equation}
To compute the integral over all the states in $\mathcal P$,
we use the Bloch-sphere representation for $\mathcal P$ given by this basis:
we parametrize the states in $\mathcal P$ using
coordinates $(\theta,\phi)$ as follows,
\begin{equation*}
\begin{split}
\ket{\psi}= \cos(\theta/2) \ket{\psi_1}+ e^{i \phi}\sin(\theta/2) \ket{\psi_2},\quad
d\psi=\frac{\sin \theta}{4\pi} d\theta d\phi
\end{split}
\end{equation*}
where $\phi\in[0,2\pi)$ and $\theta\in[0,\pi]$.

Since  $\bra{\psi} U_P^\dagger  U_P \ket{\psi}$ is independent of $\phi$ by construction, the integration over $\phi$
can be performed using the identity
$\int d\phi e^{im \phi}=2 \pi \delta_{m0}$. The result is,
\begin{equation*}
\begin{split}
F_{\mathcal P}(U) = \frac{1}{4} \int_0^\pi &\frac{d \theta \sin \theta}{a+b \sin \theta}\Bigl(
C \cos^4({\theta }/{2})
+S \sin ^4({\theta }/{2})
\\
&{}+
T \sin^2({\theta }/{2}) \cos ^2({\theta }/{2}) \Bigr)
\end{split}
\end{equation*}
where, $a$, $b$, $C$, $S$ and $T$ are the following $\theta$ independent quantities,
\begin{equation*}
\begin{split}
a &= (x_I^2 + x_R^2 + \csc^2 w)/2, \quad
b =  x_R\csc w,\\
C&=\frac{1}{2} \Bigl({x_I}^2+{x_R}^2+2-({x_I}^2+{x_R}^2)\cos (2 w) +4 {x_R} \sin w\Bigr)
,\\
S&=\frac{1}{2} \Bigl({x_I}^2+{x_R}^2+2-({x_I}^2+{x_R}^2)\cos (2 w) -4 {x_R} \sin w\Bigr)
,\\
T&=2 \Bigl(({x_I}^2+{x_R}^2)\cos (2 w) +\csc ^2w\Bigr)
.
\end{split}
\end{equation*}
This integral can be computed analytically, yielding an expression that depends on
$x_R$, $x_I$ and $w$. To obtain a result valid for an arbitrary $U_P$, we
express them using quantities invariant under rescaling and unitary
transformations of $U_P$. This can be achieved by noting that, if we define the
operator $u_P=U_P/\Tr( U_P)$, then $x_R$ is related to the norm of the
Hermitian part of $u_P$, while $x_I^2+\cot^2 w$ corresponds to the norm of its
anti-Hermitian part. The product $x_I x_R$ is given by the inner product between these
two parts.

At this stage, we also express all quantities in terms of the full operator $u_P$, rather than
restricting it to  its  $\mathcal P\mathcal P$ block component.

Thus, by working with the following quantities,
\begin{equation}
  \begin{split}
\chi&=\frac{1}{\sqrt{2}}\sqrt{\Tr[(u_P + u_P^\dagger-P)^2]},\\
\zeta &= \frac{1}{2 i } \Tr[(u_P + u_P^\dagger - P) (u_P - u_P^\dagger)],\\
\xi^2&= 3 \Bigl(1-\frac{1}{2}  \Tr[(u_P - u_P^\dagger)^2] \Bigr) ,\\
  \end{split}
\end{equation}
we can obtain,
\begin{equation}
  \begin{split}
    x_R=\chi,\,
x_I=\frac{\zeta}{\chi},\, \csc^2 w=\frac{\xi^2}{3}-\frac{\zeta^2}{\chi^2},
  \end{split}
\end{equation}
leading to the expression,
\begin{align}
F_{\mathcal P}(U)&= \frac{1}{6 A^5} \Bigl(
 A^2 (-4 A^2 + 2 A^3 + B^2 + A (B^2 + \xi^2))
\nonumber\\&{}
+(B^2 - A^2) (6 A^2 - B^2 - A (B^2 + \xi^2)) g(B/A)
 \Bigr),
\label{eq:FreezingFidelity2Plane}
\end{align}
where we defined,
\begin{equation}
  \begin{split}
A= \frac{1}{6} (\xi^2+3 \chi^2),\quad
B= \frac{1}{\sqrt{3}}{\sqrt{\xi ^2 {\chi}^2-3 \zeta ^2}},
  \end{split}
\end{equation}
and introduced the function,
\begin{equation}
  \begin{split}
g(x)=\frac{x^3+3 x-3 \operatorname{arctanh} (x)}{x^5},
  \end{split}
\end{equation}
which is regular at the origin.
\section{On the robustness of the proposed protocol}
An area that could (and should) be further explored is the robustness of the proposed protocol. Two potential sources of error, which were pointed out to us by an anonymous referee, are (i) the deviation of the Hamiltonian from the exact polynomial form assumed here and (ii) the failure to deliver the super-Zeno pulses at the exact required timing. For the first of these, note that for a specific spin $s$, the $\mathfrak{su}(2)$ generators $S_i$ obey their characteristic polynomial, of degree $2s$, so that any function of them reduces to a polynomial of maximal degree $2s$. If the anticoherence order of the plane used in the protocol is less than the polynomial degree of the perturbation, the evolution within the plane will not be suppressed optimally. Regarding possible pulse-timing errors, there is ample discussion, in the dynamical decoupling literature, of the effects of small deviations from the exact timing, $t_i \mapsto t_i +\delta t_i$~(see, \emph{e.g.}, \cite{Bie.Doh.Uys:11,Sou.Alv.Sut:12}), which may be caused by a host of experimental complications, for example, rounding errors associated with the use of digital clocking. In general, such errors degrade the performance of the protocol --- quantifying the effect requires specifying the statistics of the  deviations $\delta t_i$, in particular, their correlations, and their amplitude relative to the pulse separation. Another source of protocol degradation is the finite duration of the pulses (\emph{vs.} the instantaneous idealization assumed here). A careful study of these matters, which lies outside the scope of the present work, would help evaluate the practical relevance of the proposed protocol, and point to potential improvements.

\bibliography{referencesMQZ.bib}

\end{document}